\documentclass[12pt,draftclsnofoot,journal,onecolumn]{IEEEtran}
\usepackage{cite}
\usepackage{threeparttable}
\usepackage{graphicx}
\usepackage{picinpar}
\usepackage{color}
\usepackage[cmex10]{amsmath}
\usepackage{amsmath,amsfonts,amssymb}
\usepackage{subfigure}
\usepackage{algorithm}
\usepackage{algorithmic}
\usepackage{stfloats}
\usepackage{bm}
\usepackage{amsthm}

\begin{document}

\title{Two-Timescale Channel Estimation for Reconfigurable Intelligent Surface Aided Wireless Communications}

\author{
\IEEEauthorblockN{{Chen Hu, {\it Student Member, IEEE} and Linglong Dai, {\it Senior Member, IEEE}}}

\thanks{C. Hu and L. Dai are with the Department of Electronic Engineering, Tsinghua University, Beijing 100084, P. R. China (e-mail: huc16@mails.tsinghua.edu.cn, daill@tsinghua.edu.cn).}
\thanks{This work was supported by the National Science and Technology Major Project of China under Grant 2018ZX03001004 and the National Natural Science Foundation of China for Outstanding Young Scholars under Grant 61722109.}
}

\maketitle
\begin{abstract}
Channel estimation is challenging for the reconfigurable intelligent surface (RIS)-aided wireless communications. Since the number of coefficients of the cascaded channel among the base station (BS), the RIS and the user equipments (UEs) is the product of the number of BS antennas, the number of RIS elements, and the number of UEs, the pilot overhead can be prohibitively high. In this paper, we propose a two-timescale channel estimation framework to exploit the property that the BS-RIS channel is high-dimensional but quasi-static, while the RIS-UE channel is mobile but low-dimensional. Specifically, to estimate the quasi-static BS-RIS channel, we propose a dual-link pilot transmission scheme, where the BS transmits downlink pilots and receives uplink pilots reflected by the RIS. Then, we propose a coordinate descent-based algorithm to recover the BS-RIS channel. Since the quasi-static BS-RIS channel is estimated less frequently than the mobile channel be, the average pilot overhead can be reduced from a long-term perspective. Although the mobile RIS-UE channel has to be frequently estimated in a small timescale, the associated pilot overhead is low thanks to its low dimension. Simulation results show that the proposed two-timescale channel estimation framework can achieve accurate channel estimation with low pilot overhead.

\end{abstract}


\section{Introduction}

The emerging reconfigurable intelligent surface (RIS) has been recognized as a potential technology for the future 6G communications \cite{6GOverview}. Different from conventional wireless communications where the propagation environment between the base station (BS) and the user equipments (UEs) is considered uncontrollable, the RIS enables us to manipulate the wireless propagation environment by controlling the reflection coefficients on all RIS elements \cite{RISOverview1}. In the RIS-aided wireless communication system, accurate channel state information (CSI) is required to design the precoding matrix and the RIS reflection coefficients \cite{RISOverview2}. Consequently, it is of great importance to estimate the CSI in the RIS-aided wireless communication systems.

However, channel estimation for the RIS-aided wireless communication systems is a challenging problem due to the following two difficulties. Firstly, there are no active transmitters or receivers at the passive RIS, so the RIS can neither transmit nor receive pilots. By utilizing the active transceivers at the BS and UEs, most existing channel estimation methods only estimate the cascaded channel among the BS, the RIS and the UEs, i.e., the BS-RIS-UE cascaded channel \cite{RISCE_Rayleigh_OnOff, RISCE_Rayleigh_MVU, RISCE_structured_sparse, RISCE_Rayleigh_multiuser_reduced, RISCE_mmWave_BiGAMP, RISCE_mmWave_compressive, RISCE_CSDL, RISCE_DL, RISCE_Broadband}, where the BS-RIS-UE cascaded channel is the compound of the channel between the BS and the RIS (BS-RIS channel), and the channel between the RIS and the UEs (RIS-UE channel). Although most precoding schemes are based on the knowledge about the cascaded channel, there exist some precoding schemes \cite{RISPC_1,RISPC_2,RISPC_3} that require the individual CSI about both the BS-RIS channel and the RIS-UE channel. More importantly, the pilot overhead for the cascaded channel estimation methods is prohibitively high. A typical RIS-aided multi-user wireless communication system has a large number of BS antennas, a large number (e.g., dozens to hundreds) of RIS elements and dozens of UEs, while the pilot overhead of typical cascaded channel methods is the product of the number of RIS elements and the number of UEs \cite{RISCE_Rayleigh_OnOff,RISCE_Rayleigh_MVU}. Consequently, the pilot overhead to estimate the BS-RIS-UE cascaded channel can be prohibitively high (e.g., hundreds to thousands) in practice.

\vspace{3mm}
{\noindent \it A. Prior works}

Up to now, the research on channel estimation for the RIS-aided communication systems is limited. There are only a few recent works on this topic \cite{RISCE_Rayleigh_OnOff, RISCE_Rayleigh_MVU, RISCE_Rayleigh_multiuser_reduced, RISCE_mmWave_compressive, RISCE_mmWave_BiGAMP, RISCE_structured_sparse,RISCE_CSDL,RISCE_Broadband,RISCE_DL}, all of which estimate the BS-RIS-UE cascaded channel. In \cite{RISCE_Rayleigh_OnOff}, only a single RIS element is turned on in one time slot, and the cascaded channel of this RIS element can be estimated. The channel estimation accuracy suffers from the degradation of the receiver signal-to-noise ratio (SNR), because all other RIS elements do not reflect the pilots. The authors in \cite{RISCE_Rayleigh_MVU} designed a series of reflection coefficient vectors, and achieved a minimum variance unbiased (MVU) estimate of the cascaded channel. However, the pilot overhead for these schemes \cite{RISCE_Rayleigh_OnOff, RISCE_Rayleigh_MVU} equals to the number of RIS elements multiplied by the number of UEs, which prohibits their application in a system with a large number of RIS elements. In \cite{RISCE_Rayleigh_multiuser_reduced}, the authors propose an ingenious method to reduce the pilot overhead. The cascaded channel of the first UE is estimated at first and used as a reference channel. Then the cascaded channel of other UEs are estimated with reduced pilot overhead. However, the estimation is not accurate when the SNR is low. Another solution to reduce the pilot overhead is to use the sparse matrix factorization and matrix completion method if the channel exhibits the low-rank property \cite{RISCE_mmWave_BiGAMP}. In addition, the spatial channel sparsity can be leveraged to reduce the pilot overhead based on the compressive sensing technique for the RIS-aided communication systems operating at high-frequency bands \cite{RISCE_mmWave_compressive, RISCE_structured_sparse}. \cite{RISCE_CSDL,RISCE_DL} use deep learning tools to solve the problem. The wideband channel estimation is discussed in \cite{RISCE_Broadband}. However, these methods are not applicable in the general scenario where the channel is not low-rank or sparse.

\vspace{3mm}
{\noindent \it B. Contributions}

In this paper, by leveraging the two-timescale property of the channel, we propose a two-timescale channel estimation framework to significantly reduce the pilot overhead in RIS-aided wireless communication systems. The contributions are summarized as follows:

\begin{enumerate}

\item We exploit the two-timescale channel property that the BS-RIS channel is quasi-static since the BS and the RIS are fixed, while the RIS-UE channel and the the channel between the BS and the UEs (BS-UE channel) can change with time due to the mobility of the UEs. Then, we propose a two-timescale channel estimation framework to significantly reduce the pilot overhead. In this framework, the quasi-static BS-RIS with high dimension is estimated in a large timescale, while the mobile RIS-UE and BS-UE channels with low dimension are estimated in a small timescale.

\item  To estimate the quasi-static BS-RIS channel, we propose a dual-link pilot transmission scheme, where the BS transmits downlink pilots and receives uplink pilots reflected by the RIS. Then, we propose a coordinate descent-based algorithm to recover the BS-RIS channel. Since the quasi-static BS-RIS channel is estimated less frequently than the mobile channel estimation for the UEs, the average pilot overhead associated to the former stage can be reduced from a long-term perspective.

\item For the mobile RIS-UE and the BS-UE channels, they can be easily estimated by the existing solutions such as the least square algorithm. Although they have to be frequently estimated in a small timescale, their dimension is smaller than that of the BS-RIS-UE cascaded channel, which still has to be frequently estimated in the existing cascaded channel estimation methods, so the associated pilot overhead can be significantly reduced.

\end{enumerate}

\vspace{3mm}
{\noindent \it C. Organization and Notations}

The remainder of the paper is organized as follows. In Section II, we introduce the model of the RIS-aided wireless communication systems. In Section III, we propose the two-timescale channel estimation framework, to exploit the property that the BS-RIS channel is high-dimensional but quasi-static, while the RIS-UE and BS-UE channels are mobile but low-dimensional. The proposed large-scale channel estimation method and the small-timescale channel estimation method is discussed in Section IV and Section V, respectively. Then, the pilot overhead and the computational complexity are analyzed in Section VI. We provide the simulation results in Section VII, and conclude this paper in Section VIII.

\emph{Notations: }In this paper, light symbols, boldface lower-case symbols and upper-case symbols denote scalars, column vectors and matrices, respectively. $\left(\cdot\right)^*$ and $\left|\cdot\right|$ denote the conjugate and the amplitude, while $\left\|\cdot\right\|_0$, $\left\|\cdot\right\|_2$ and $\left\|\cdot\right\|_{F}$ denote the $\ell_{0}$-norm, the $\ell_{2}$-norm and the Frobenius norm, respectively. $(\cdot)^{T}$ and $(\cdot)^{H}$ are the transpose and the conjugate transpose, respectively. $\lceil x\rceil$ denotes the smallest integer that is greater than or equal to $x$. ${\rm{diag}}({\bf{x}})$ is the diagonal matrix with the vector ${\bf{x}}$ on its diagonal. ${\rm{vec}}({\bf{X}})$ is the vectorization of the matrix $\bf X$. $\odot$ is the Hadamard product. ${\bf I}_X$ is the $X\times X$ identity matrix. Finally, ${\mathcal {CN}}\left({\boldsymbol \mu},{\bf \Sigma}\right)$ is the complex-valued Gaussian distribution with mean ${\boldsymbol \mu}$ and covariance ${\bf \Sigma}$.

\section{System Model}\label{S1}

\begin{figure}[tp!]
\begin{center}
\includegraphics[height=0.6\columnwidth, angle=270]{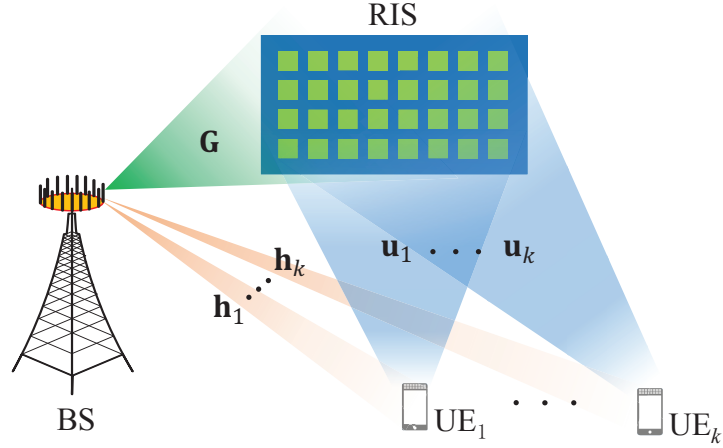}
\end{center}
\caption{The RIS aided wireless communication system.}
\label{fig1}
\end{figure}

We consider an RIS-aided wireless communication system as shown in Fig. 1. $K$ UEs are served simultaneously by a BS with $M$ antennas and an RIS with $N$ elements. The uplink signal model is given by \cite{RISOverview2}
\begin{equation}\label{model:RIS_signal_model}
{\bf y} = \sum\limits_{k=1}^{K}\left[{\bf G}\left({\boldsymbol \phi}\odot{\bf f}_k\right)+{\bf h}_k \right]x_k + {\bf n},
\end{equation}
where ${\bf y} \in {\mathbb C}^{M\times 1}$ is the received signal at the BS, ${\bf G} \in {\mathbb C}^{M \times N}$ is the BS-RIS channel, ${\boldsymbol \phi} \in {\mathbb C}^{N\times 1}$ is the vector of reflection coefficients on the RIS elements, ${\bf f}_k \in {\mathbb C}^{N\times 1}$ is the channel between the RIS and the $k$-th UE, ${\bf h}_k \in {\mathbb C}^{M\times 1}$ is the direct channel between the BS and the $k$-th UE, $x_k \in {\mathbb C}$ is the transmitted signal from the $k$-th UE, and ${\bf n} \in {\mathbb C}^{M\times 1}$, ${\bf n}\sim {\mathcal {CN}}\left({\bf 0},\sigma^2_n{\bf I}_M\right)$ is the additive noise. We assume uncorrelated Rayleigh fading channels, i.e., ${\bf h}_k \sim {\mathcal {CN}}\left({\bf 0},\rho_{h_k}{\bf I}_M\right)$, ${\rm vec}\left({\bf G}\right) \sim {\mathcal {CN}}\left({\bf 0},\rho_{g}{\bf I}_{MN}\right)$, ${\bf f}_k \sim {\mathcal {CN}}\left({\bf 0},\rho_{f_k}{\bf I}_N\right)$, where $\rho_{h_k}$, $\rho_{g}$ and $\rho_{f_k}$ are the large-scale fading factors of the three channels, respectively. As demonstrated in the existing RIS hardware prototype \cite{RIS_prototype}, RIS elements are mainly used to control the phases of the reflected signal, i.e., $\left|{\boldsymbol \phi}\left(n\right)\right|=1$, $1\le n\le N$. The signal model (\ref{model:RIS_signal_model}) can be equivalently written as
\begin{equation}\label{model:RIS_signal_model_equivalent}
\begin{aligned}
{\bf y} &= \sum\limits_{k=1}^{K}\left[{\bf G}{\rm diag}\left({\bf f}_k\right){\boldsymbol \phi}+{\bf h}_k\right]x_k + {\bf n}\\
&=\sum\limits_{k=1}^{K}\left[{\bf C}_k{\boldsymbol \phi}+{\bf h}_k\right]x_k + {\bf n},
\end{aligned}
\end{equation}
where the BS-RIS-UE cascaded channel is defined by
\begin{equation}\label{equ:definition_of_cascaded_channel}
{\bf C}_k\triangleq {\bf G}{\rm diag}\left({\bf f}_k\right),~~k=1,2,\cdots,K,
\end{equation}
which is the compound of the BS-RIS channel and the RIS-UE channel \cite{RISCE_mmWave_BiGAMP}.

Most cascaded channel estimation methods in literature are based on the uplink pilot transmission \cite{RISCE_Rayleigh_MVU, RISCE_structured_sparse, RISCE_Rayleigh_multiuser_reduced, RISCE_mmWave_BiGAMP}. In the $t$-th time slot, different UEs transmit different pilots $x_{k,t}$, while the RIS use the same reflection coefficient vector ${\boldsymbol \phi}_t$ to reflect the uplink pilots from all UEs. The received pilots at the BS can be modeled by
\begin{equation}\label{model:uplink_pilot_transmission}
\begin{aligned}
{\bf y}_t &= \sum\limits_{k=1}^{K}\left[{\bf C}_k{\boldsymbol \phi}_t + {\bf h}_k\right]x_{k,t} + {\bf n}_t\\
&=\sum\limits_{k=1}^{K}{\bf C}_k{\boldsymbol \phi}_t x_{k,t} + \sum\limits_{k=1}^{K}{\bf h}_k x_{k,t}  + {\bf n}_t.
\end{aligned}
\end{equation}
We repeat (\ref{model:uplink_pilot_transmission}) for multiple time slots to received enough pilots, then the BS-RIS-UE cascaded channel $\left\{{\bf C}_k|1\le k\le K\right\}$ and the BS-UE channel $\left\{{\bf h}_k|1\le k\le K\right\}$ can be directly estimated by the methods in \cite{RISCE_Rayleigh_OnOff, RISCE_Rayleigh_MVU}. Since the dimension of the received pilots should be no smaller than the dimension of the channels in \cite{RISCE_Rayleigh_MVU, RISCE_Rayleigh_OnOff}, the pilot overhead is extremely large to estimate the $\left(MNK+MK\right)$ coefficients in the BS-RIS-UE cascaded channel and the BS-UE direct channel for all the UEs.

\section{The Proposed Two-timescale Channel Estimation Framework}

In this section, we propose a two-timescale channel estimation framework to leverage the two-timescale channel property. On one hand, the BS and the RIS are placed in fixed positions, so the BS-RIS channel $\bf G$ is quasi-static. We only need to estimate $\bf G$ in a large timescale, i.e., estimate $\bf G$ once over a long period of time. On the other hand, the RIS-UE channel and the BS-UE channel are time-varying due to the mobility of the UEs, so we need to estimate them in a small timescale, i.e., estimate them once in a short period of time.

To estimate the quasi-static BS-RIS channel, the main difficulty is that the RIS can neither transmit nor receive pilots because the RIS does not have active transceivers \cite{RISOverview1}. To overcome this difficulty, we propose a dual-link pilot transmission scheme. Specifically, the BS works at full-duplex mode \cite{FD_SI_Cancellation1}. The BS transmits pilots to the RIS via the downlink channel with a single antenna, and then the RIS reflects pilots back to the BS via the uplink channel with a set of pre-designed reflection coefficients, which will be explained later in detail. At the same time the BS also receives pilots with the rest antennas. Though the self-interference can be severe in the full-duplex system, the self-interference mitigation techniques have been extensively studied to solve this problem, e.g., \cite{FD_SI_Cancellation1, FD_SI_Cancellation2, FD_SI_Cancellation3}. After the self-interference mitigation, the BS-RIS channel can be estimated based on the dual-link pilots which are received at the BS. Then, for the mobile RIS-UE and BS-UE channels, since they are low-dimensional, they can be estimated with a conventional uplink pilot transmission scheme and an LS-based algorithm.

Compared with the existing cascaded channel estimation methods, the pilot overhead can be significantly reduced. On one hand, the BS-RIS channel is high-dimensional but quasi-static. Since the quasi-static BS-RIS channel is estimated less frequently than the mobile channel estimation for the UEs, the average pilot overhead associated to the former stage can be reduced from a long-term perspective. On the other hand, the BS-UE and RIS-UE channels are mobile but low-dimensional. We estimate $\left\{{\bf h}_k,{\bf f}_k|1\le k\le K\right\}$ with only $\left(M+N\right)K$ coefficients, rather than estimating $\left\{{\bf h}_k,{\bf C}_k|1\le k\le K\right\}$ with $\left(M+MN\right)K$ coefficients by the cascaded channel estimation methods. The required pilot overhead can thus be significantly reduced.

The proposed channel estimation frame structure is exhibited in Fig. 2. At the beginning, we estimate the high-dimensional quasi-static BS-RIS channel based on the proposed dual-link pilot transmission scheme. Then, in a small timescale, the low-dimensional mobile BS-UE and RIS UE channels are estimated based on the uplink pilots before data transmission.

\begin{figure}[tp!]
\begin{center}
\includegraphics[width=0.7\columnwidth]{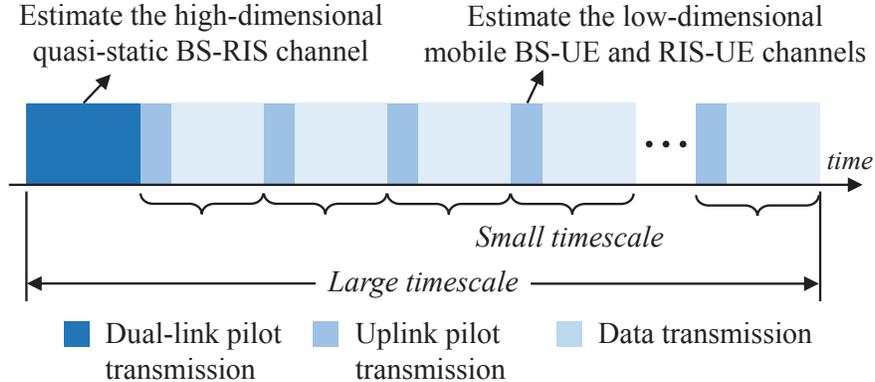}
\end{center}
\caption{The proposed two-timescale channel estimation frame structure.}
\label{fig1}
\end{figure}

\section{Quasi-static BS-RIS Channel Estimation}

In this section, we propose a dual-link pilot transmission scheme and a coordinate descent-based BS-RIS channel estimation algorithm to estimate the quasi-static BS-RIS channel.

\subsection{Dual-link pilot transmission}$~$

Before we start, we show that we cannot uniquely estimate the BS-RIS channel based on the pilot transmission model in (\ref{model:uplink_pilot_transmission}), which is why we propose a novel dual-link pilot transmission scheme. Note that according to the definition of the cascaded channel in (\ref{equ:definition_of_cascaded_channel}), for any non-zero $p_1,\cdots,p_{N}\in {\mathbb C}$,
\begin{equation}\label{equ:decomposition_ambiguity}
\begin{aligned}
{\bf C}_k&={\bf G}{\rm diag}\left({\bf f}_k\right)\\
&=\left({\bf G}\left[\begin{matrix}p_1 & & \\ & \ddots & \\ & & p_{N}\end{matrix}\right]\right)\left(\left[\begin{matrix}p_1^{-1} & & \\ &  \ddots & \\ & & p^{-1}_{N}\end{matrix}\right]{\rm diag}\left({\bf f}_k\right)\right)\\
&={\bf G}'{\rm diag}\left({\bf f}_k'\right),~~k=1,2,\cdots,K,
\end{aligned}
\end{equation}
where ${\bf G}'={\bf G}{\rm diag}\left(\left[p_1,\cdots,p_{N}\right]^T\right)$, ${\bf f}_k'={\bf f}_k\odot \left[p_1,\cdots,p_{N}\right]^T$. Equation (\ref{equ:decomposition_ambiguity}) shows that the decomposition of the cascaded channel ${\bf C}_k$ is not unique. Therefore, no matter what pilots and no matter what reflection coefficients are used in the uplink pilot transmission in (\ref{model:uplink_pilot_transmission}), both ${\bf G},\left\{{\bf f}_k|1\le k\le {K}\right\}$ and ${\bf G}',\left\{{\bf f}_k'|1\le k\le K\right\}$ can lead to the same received pilots. As a result, we cannot uniquely estimate ${\bf G}$ and $\left\{{\bf f}_k|1\le k\le K\right\}$ based on the conventional uplink pilot transmission model in (\ref{model:uplink_pilot_transmission}) \cite{RISCE_mmWave_BiGAMP}. So, we have to propose a pilot transmission scheme different from (\ref{model:uplink_pilot_transmission}) to estimate the quasi-static BS-RIS channel.

In the proposed dual-link pilot transmission scheme, we do not need the UEs to transmit or receive pilots. The key idea of the dual-link pilot transmission is that the BS transmits pilots to the RIS via the downlink channel, and then the RIS reflects pilots back to the BS via the uplink channel. Assuming we have a full-duplex BS\footnote{The full-duplex BS is one of the possible realizations of the dual-link pilot transmission. Other realizations remain open problems for future works.} that can transmit and receive pilots with different antennas simultaneously\cite{FD_SI_Cancellation1}.

To be specific, the proposed dual-link pilot transmission frame consists of $\left(N+1\right)$ sub-frames, and each sub-frame lasts for $L$ time slots. In the $t$-th ($t=1,2,\cdots,N+1$) sub-frame, the reflection coefficient vector at the RIS is ${\bar {\boldsymbol \phi}}_t \in {\mathbb C}^{N\times 1}$. In the $m_1$-th ($m_1=1,2,\cdots,L$) time slot of the $t$-th sub-frame, the $m_1$-th BS antenna transmits a pilot $z_{m_1,t}$, and the rest $\left(M-1\right)$ BS antennas receive the pilot reflected by the RIS. The received pilots at the rest BS antennas can be written as
\begin{equation}\label{model:pilot_model_scalar}
\begin{aligned}
\!{\bar y}_{m_1\!,m_2\!,t} \!&=\!\! \left[{\bf g}_{m_2}^T\!{\rm diag}\!\left({\bar {\boldsymbol \phi}}_t\right)\!{\bf g}_{m_1} \!\!+\!\! s_{m_1\!,m_2}\right]\!\!z_{m_1\!,t} \!+\! {\bar i}_{m_1\!,m_2\!,t} \!+\! {\bar n}_{m_1\!,m_2\!,t}\\
&= \!\!\left[\left({\bf g}_{m_1}\!\odot\!{\bf g}_{m_2}\right)^T\!\!{\bar {\boldsymbol \phi}}_t \!+ \!\! s_{m_1\!,m_2}\right]\!\!z_{m_1\!,t} \!+\! {\bar i}_{m_1\!,m_2\!,t} \!+\! {\bar n}_{m_1\!,m_2\!,t},\\
&~~~~~~m_2=1,2,\cdots,M, m_2\neq m_1,
\end{aligned}
\end{equation}
where $m_1\neq m_2$ means that the receiver antenna is different from the transmitter antenna. ${\bar y}_{m_1,m_2,t}\in {\mathbb C}$ is the received pilot at the $m_2$-th BS antenna, ${\bf g}_{m_2}^T \triangleq {\bf G}\left(m_2,:\right) \in {\mathbb C}^{N\times 1}$, ${\bf g}_{m_1} \triangleq {\bf G}\left(m_1,:\right)^T\in {\mathbb C}^{N\times 1}$. Since $\bf G$ is BS-RIS channel matrix, ${\bf g}_{m_2}^T$, ${\bf g}_{m_1}^T$ are row vectors that denote the uplink channels from the RIS to the $m_2$-th and to the $m_1$-th BS antenna, respectively. Thanks to the channel reciprocity, the column vector ${\bf g}_{m_1}$ can be used to express the downlink channel from the $m_1$-th BS antenna to the RIS. The term $s_{m_1,m_2}$ is used to represent the environmental reflection, since some objects other than the RIS elements can also reflect the pilot sent by the $m_1$-th BS antenna to the $m_2$-th RIS antenna. ${\bar i}_{m_1,m_2,t}$ is the self-interference after mitigation, which will be explained later. Finally, ${\bar n}_{m_1,m_2,t}\sim {\mathcal {CN}}\left(0,\sigma_n^2\right)$ is the received noise at the $m_2$-th BS antenna.

The self-interference ${\bar i}_{m_1,m_2,t}$ is mainly caused by the direct transmission from the $m_1$-th BS antenna to the $m_2$-th BS antenna when the BS is working at the full-duplex mode. In a typical full-duplex system, the self-interference can be even larger than the desired signal (e.g., 110 dB larger than the desired signal \cite{FD_SI_Cancellation1}) before mitigation. However, the self-interference suppression methods have been extensively studied to solve the problem, e.g., \cite{FD_SI_Cancellation1,FD_SI_Cancellation2,FD_SI_Cancellation3}. In \cite{FD_SI_Cancellation1}, the authors reported that the interference can be mitigated to as low as about only 3~dB higher than the receiver noise with their self-interference mitigation method. Therefore, we assume ${\bar i}_{m_1,m_2,t}\sim {\mathcal {CN}}\left(0,\sigma_{i}^2\right)$ after the self-interference mitigation, where there is no considerable differences in orders of magnitude between $\sigma_{i}^2$ and $\sigma_n^2$. It is true that the self-interference suppression will result in increased hardware complexity at the BS. It is still an open topic to study the RIS-aided full-duplex system in the future.

Denote the index tuple set ${\mathcal S}\triangleq\left\{\left(m_1,m_2\right)|1\le m_1\le L,\right.$ $\left. 1\le m_2\le M,m_1\neq m_2\right\}$. After $\left(N+1\right)$ sub-frames, we can get all the received pilots $\left\{{\bar y}_{m_1,m_2,t}|\left(m_1,m_2\right)\in {\mathcal S},\right.$ $\left.1\le t\le N+1\right\}$ in the dual-link pilot transmission frame. For a given $\left(m_1,m_2\right)\in {\mathcal S}$, we collect the received pilots corresponding to the $m_1$-th transmit antenna and the $m_2$-th receive antenna throughout $\left(N+1\right)$ sub-frames, to define
\begin{equation}\label{equ:collected_y_vector}
{\bar {\bf y}}_{m_1,m_2}^T \triangleq \left[{\bar y}_{m_1,m_2,1},{\bar y}_{m_1,m_2,2},\cdots,{\bar y}_{m_1,m_2,N+1}\right].
\end{equation}
Then, by substituting the dual-link pilot model (\ref{model:pilot_model_scalar}) into (\ref{equ:collected_y_vector}), and assuming the transmitted pilot is $z_{m_1,t}=\sqrt{P_{\rm BS}}$ without loss of generality, where $P_{\rm BS}$ is the transmitted power of the BS, we can write the model in the vector form:
\begin{equation}\label{model:collected_model}
\begin{aligned}
{\bar {\bf y}}_{m_1,m_2}^T &=\left\{\left({\bf g}_{m_1}\odot{\bf g}_{m_2}\right)^T\left[{\bar {\boldsymbol \phi}}_1,{\bar {\boldsymbol \phi}}_2,\cdots,{\bar {\boldsymbol \phi}}_{N+1}\right]\right. \\
&~~~\left.+ s_{m_1,m_2}{\bf 1}_{1 \times \left(N+1\right)}\right\}\sqrt{P_{\rm BS}} + {\bar {\bf i}}_{m_1,m_2}^T+{\bar {\bf n}}_{m_1,m_2}^T\\
&=\sqrt{P_{\rm BS}}{\bf w}_{m_1,m_2}^T\left[ \begin{matrix} {\bf 1}_{1 \times \left(N+1\right)}\\ {\bar {\bf \Phi}}\end{matrix} \right] + {\bar {\bf i}}_{m_1,m_2}^T+ {\bar {\bf n}}_{m_1,m_2}^T,
\end{aligned}
\end{equation}
where we define the vector of unknown variables by
\begin{equation}\label{equ:definition_of_w}
{\bf w}_{m_{1},m_{2}}\triangleq \left[s_{m_{1},m_{2}}~~ \left({\bf g}_{m_1}\odot{\bf g}_{m_2}\right)^T\right]^T,
\end{equation}
and ${\bar {\bf \Phi}} = \left[{\bar {\boldsymbol \phi}}_1,{\bar {\boldsymbol \phi}}_2,\cdots,{\bar {\boldsymbol \phi}}_{N+1}\right]\in {\mathbb C}^{N \times \left({N+1}\right)}$, ${\bf 1}_{1 \times \left(N+1\right)}$ is a row vector with all elements equal to 1, ${\bar {\bf i}}_{m_1,m_2} = \left[{\bar i}_{m_1,m_2,1},\cdots,{\bar i}_{m_1,m_2,N+1}\right]^T$, ${\bar {\bf n}}_{m_1,m_2} = \left[{\bar n}_{m_1,m_2,1},\cdots,\right.$ $\left.{\bar n}_{m_1,m_2,N+1}\right]^T$.

For the convenience of signal processing, the reflection coefficient vectors are designed as
\begin{equation}
\begin{aligned}
{\bar {\boldsymbol \phi}}_t &= \left[e^{-j2\pi\frac{1\left(t-1\right)}{N+1}},e^{-j2\pi\frac{2\left(t-1\right)}{N+1}},\cdots,e^{-j2\pi\frac{N\left(t-1\right)}{N+1}}\right]^T,\\
&~~~~~~~~~~~~~~~~~t=1,2,\cdots,N+1,
\end{aligned}
\end{equation}
which leads to
\begin{equation}
\begin{aligned}
\left[ \begin{matrix} {\bf 1}_{1 \times \left(N+1\right)}\\ {\bar {\bf \Phi}}\end{matrix} \right]&=\left[ \begin{matrix} 1 & 1 & \cdots & 1 \\ 1 & e^{-j2\pi\frac{1}{N+1}} & \cdots & e^{-j2\pi\frac{N}{N+1}} \\ \vdots & \vdots & \ddots & \vdots \\ 1 & e^{-j2\pi\frac{N}{N+1}} & \cdots & e^{-j2\pi\frac{N^2}{N+1}}\end{matrix} \right]\\&=\sqrt{N+1}{\bf F}_{N+1},
\end{aligned}
\end{equation}
where ${\bf F}_{N+1}$ is the ${\left({N+1}\right) \times \left({N+1}\right)}$-dimensional unitary DFT matrix. Therefore, (\ref{model:collected_model}) becomes
\begin{equation}\label{model:collected_model_1}
{\bar {\bf y}}_{m_1,m_2}^T=\sqrt{\left(N\!+\!1\right)P_{\rm BS}}{\bf w}_{m_1,m_2}^T{\bf F}_{N+1} + {\bar {\bf i}}_{m_1,m_2}^T+ {\bar {\bf n}}_{m_1,m_2}^T.
\end{equation}

\subsection{Proposed coordinate descent-based channel estimation algorithm}$~$

Given the received dual-link pilots, we estimate the quasi-static BS-RIS channel in three stages. Firstly, we divide the problem into $N$ independent subproblems, where each subproblem is to estimate the channel between the BS and a single RIS element. Secondly, initial estimates are calculated for each subproblem. Thirdly, the BS-RIS channel estimates are iteratively optimized via a coordinate descent approach.

\subsubsection{Problem division}
In the first stage, for all $\left(m_1,m_2\right)\in {\mathcal S}$, based on the model (\ref{model:collected_model_1}), we can easily obtain an estimate of ${\bf w}_{m_1,m_2}$ defined in (\ref{equ:definition_of_w}) by
\begin{equation}\label{equ:estimate_w}
\begin{aligned}
{\hat {\bf w}}_{m_1,m_2}^T&\triangleq \left[{\hat s}_{m_1,m_2}~~a_{m_1,m_2,1}~~a_{m_1,m_2,2}~~\cdots~~a_{m_1,m_2,N}\right]\\
&=\frac{1}{\sqrt{\left(N+1\right)P_{\rm BS}}}{\bar {\bf y}}_{m_1,m_2}^T{\bf F}_{N+1}^H,
\end{aligned}
\end{equation}
where  ${\hat s}_{m_1,m_2}$, $a_{m_1,m_2,1}$, $\cdots$, $a_{m_1,m_2,N}$ denote the elements of ${\hat {\bf w}}_{m_1,m_2}$.

By substituting (\ref{model:collected_model}) into (\ref{equ:estimate_w}), we have
\begin{equation}
\begin{aligned}
 &\left[{\hat s}_{m_1,m_2}~~a_{m_1,m_2,1}~~a_{m_1,m_2,2}~~\cdots~~a_{m_1,m_2,N}\right]\\
&=\left[s_{m_{1},m_{2}}~~ \left({\bf g}_{m_1}\odot{\bf g}_{m_2}\right)^T\right]+\frac{\left({\bar {\bf i}}_{m_1,m_2}^T+{\bar {\bf n}}_{m_1,m_2}^T\right){\bf F}_{N+1}^H}{\sqrt{P_{\rm BS}\left(N+1\right)}}.
\end{aligned}
\end{equation}
Therefore, ${a}_{m_1,m_2,n}$ is an estimate of ${g}_{m_1,n}{g}_{m_2,n}$, with
\begin{equation}\label{model:definition_of_a}
\begin{aligned}
{a}_{m_1,m_2,n} &= {g}_{m_1,n}{g}_{m_2,n} + \varepsilon_{m_1,m_2,n},\\
n&=1,2,\cdots,N,
\end{aligned}
\end{equation}
where $g_{m,n}={\bf g}_{m}\left(n\right)={\bf G}\left(m,n\right)$ is the channel between the $m$-th BS antenna and the $n$-th RIS element, $\varepsilon_{m_1,m_2,n}$ is the error, and $\varepsilon_{m_1,m_2,n}\sim {\mathcal {CN}}\left(0,\frac{\sigma_i^2+\sigma_n^2}{P_{\rm BS}\left(N+1\right)}\right)$.

Now that for a specific $n$, $1\le n\le N$, the variables $\left\{{a}_{m_1,m_2,n}|\left(m_1,m_2\right)\in {\mathcal S}\right\}$ are dependent to the channel coefficients between the BS and the $n$-th RIS element, i.e., $\left\{{g}_{m,n}|1\le m\le M\right\}$, but independent to the channel coefficients related to other RIS elements. That is to say, the problem of estimating the quasi-static BS-RIS channel can thus be divided into $N$ independent subproblems, where the $n$-th subproblem is to estimate the channel coefficients related to the $n$-th RIS element $\left\{{g}_{m,n}|1\le m\le M\right\}$ from $\left\{{a}_{m_1,m_2,n}|\left(m_1,m_2\right)\in {\mathcal S}\right\}$. Since $\left|{\mathcal S}\right|\ge M$ is required to guarantee the solvability of the subproblem, we need to have $L\ge 2$.

In the following stages, we will estimate the quasi-static BS-RIS channel by solving the $N$ subproblems respectively. For a given $n$, we formulate the $n$-th subproblem as
\begin{equation}\label{problem:multivariate_optimization}
{\hat g}_{1,n},{\hat g}_{2,n},\cdots,{\hat g}_{M,n}=\arg\min\limits_{g_{1,n},\cdots,g_{M,n}}J_n\left(g_{1,n},\cdots,g_{M,n}\right),
\end{equation}
where
\begin{equation}\label{equ:objective_function}
J_n\left(g_{1,n},\cdots,g_{M,n}\right)\triangleq\sum\limits_{\left(m_1,m_2\right)\in {\mathcal S}}\left|a_{m_1,m_2,n}-g_{m_1,n}g_{m_2,n}\right|^2.
\end{equation}

\subsubsection{Calculate initial channel estimates}
The goal of the second stage is to calculate initial channel estimates for the subproblems in (\ref{problem:multivariate_optimization}). The initial estimates are coarse and easy to calculate, which will help the iterative refinement in the third stage converge faster. Specifically, we pick $1\le m_1<m_2\le L$ and $m_3\neq m_1,m_2$. The initial estimates ${\hat g}^{(0)}_{1,n},{\hat g}^{(0)}_{2,n},\cdots,{\hat g}^{(0)}_{M,n}$ for the $n$-th subproblem are calculated by
\begin{equation}\label{equ:init_1}
{\hat g}^{(0)}_{m_1,n}\leftarrow \sqrt{\frac{a_{m_1,m_2,n}a_{m_1,m_3,n}}{a_{m_2,m_3,n}}},
\end{equation}
\begin{equation}\label{equ:init_2}
{\hat g}^{(0)}_{m',n}\leftarrow \frac{a_{m_1,m',n}}{{\hat g}^{(0)}_{m_1,n}},~1\le m'\le M,m'\neq m_1.
\end{equation}

\subsubsection{Coordinate descent-based iterative refinement}

In the third stage, we propose a coordinate descent-based algorithm to refine the coarse channel estimates and find the accurate estimates of the quasi-static BS-RIS channel, to solve the subproblems in (\ref{problem:multivariate_optimization}). There are multiple outer iterations in the coordinate descent based algorithm, each of which consists of $M$ inner iterations. The key idea is that, in each outer iteration, the estimates of all the $M$ coefficients are refined from the first one to the last one. In each inner iteration, we refine the estimate of one of the $M$ coefficients while fixing the estimates of the rest $\left(M-1\right)$ coefficients.

To be specific, in the $i$-th outer iteration, we loop for $1\le m\le M$. In the $m$-th inner iteration, we refine the estimate of $g_{m,n}$ but fix the estimate of other $M-1$ channel coefficients at $g_{1,n}^{(i,m)},\cdots,g_{m-1,n}^{(i,m)},g_{m+1,n}^{(i,m)},\cdots,g_{M,n}^{(i,m)}$,
where
\begin{equation}
g_{m',n}^{(i,m)}\triangleq\left\{\begin{matrix} g_{m',n}^{(i)}, & m'<m,\\ g_{m',n}^{(i-1)}, & m'>m,\end{matrix}\right.
\end{equation}
which means that ${\hat g}^{(i)}_{1,n},\cdots,{\hat g}^{(i)}_{m-1,n}$ have been refined before refining ${\hat g}_{m,n}$ in the current outer iteration, and ${\hat g}^{(i-1)}_{m+1,n},\cdots,{\hat g}^{(i-1)}_{M,n}$ have been refined in the $\left(i-1\right)$-th outer iteration. Using these notations, the inner-iteration problem of refining the estimate of $g_{m,n}$ is formulated as
\begin{equation}\label{problem:univariate_optimization}
{\hat g}^{(i)}_{m,n}\!\!=\!\arg\!\min\limits_{g_{m,n}}J_n\!\!\left(\!g_{1,n}^{(i,m)}\!,\cdots\!,g_{m\!-\!1,n}^{(i,m)},g_{m,n},g_{m\!+\!1,n}^{(i,m)},\cdots\!,g_{M,n}^{(i,m)}\!\right).
\end{equation}

As a univariate optimization problem, the close-form solution for (\ref{problem:univariate_optimization}) can be derived by solving $\frac{\partial f_n}{\partial g_{m,n}}=\frac{\partial f_n}{\partial g^*_{m,n}}=0$. According to the definition of the objective function $f_n$ in (\ref{equ:objective_function}), the partial derivatives are given by
\begin{equation}
\begin{aligned}
\frac{\partial J_n}{\partial g_{m,n}}&=\frac{\partial}{\partial g_{m,n}}\sum\limits_{\left(m,m'\right)\in {\mathcal S}}\left|a_{m,m',n}-g_{m,n}g_{m',n}^{(i,m)}\right|^2\\
&~~+\frac{\partial}{\partial g_{m,n}}\sum\limits_{\left(m',m\right)\in {\mathcal S}}\left|a_{m',m,n}-g_{m',n}^{(i,m)}g_{m,n}\right|^2\\
&=g_{m,n}^*\left(\sum\limits_{\left(m,m'\right)\in {\mathcal S}}\left|g_{m',n}^{(i,m)}\right|^2+\sum\limits_{\left(m',m\right)\in {\mathcal S}}\left|g_{m',n}^{(i,m)}\right|^2\right)\\
&~~+\!\sum\limits_{\left(m,m'\right)\in {\mathcal S}}a_{m,m',n}^*g_{m',n}^{(i,m)}+\!\sum\limits_{\left(m',m\right)\in {\mathcal S}}a_{m',m,n}^*g_{m',n}^{(i,m)},
\end{aligned}
\end{equation}
and similarly
\begin{equation}
\begin{aligned}
\frac{\partial J_n}{\partial g_{m,n}^*}&=g_{m,n}\left(\sum\limits_{\left(m,m'\right)\in {\mathcal S}}\left|g_{m',n}^{(i,m)}\right|^2+\sum\limits_{\left(m',m\right)\in {\mathcal S}}\left|g_{m',n}^{(i,m)}\right|^2\right)\\
&\!\!\!+\!\!\!\sum\limits_{\left(m,m'\right)\in {\mathcal S}}\!\!a_{m,m'\!,n}\!\left(g_{m',n}^{(i,m)}\right)^*\!+\!\!\!\sum\limits_{\left(m',m\right)\in {\mathcal S}}\!\!a_{m'\!,m,n}\!\left(g_{m',n}^{(i,m)}\right)^*.
\end{aligned}
\end{equation}

Therefore, the close-form solution to (\ref{problem:univariate_optimization}) is obtained by $\frac{\partial f_n}{\partial g_{m,n}}=\frac{\partial f_n}{\partial g^*_{m,n}}=0$, which yields to
\begin{equation}\label{equ:update_gmn_est}
{\hat g}_{m,n}^{(i)}=\frac{\!\!\sum\limits_{\left(m,m'\right)\in {\mathcal S}}\!\!a_{m,m',n}\!\left(\!g_{m',n}^{(i,m)}\!\right)^*\!\!+\!\!\sum\limits_{\left(m',m\right)\in {\mathcal S}}\!\!a_{m',m,n}\!\left(\!g_{m',n}^{(i,m)}\!\right)^*}{\sum\limits_{\left(m,m'\right)\in {\mathcal S}}\left|g_{m',n}^{(i,m)}\right|^2+\sum\limits_{\left(m',m\right)\in {\mathcal S}}\left|g_{m',n}^{(i,m)}\right|^2}.
\end{equation}


The coordinate descent-based channel estimation algorithm is summarized in {\bf Algorithm 1}. In Steps 1-6, we divide the quasi-static channel estimation problem into $N$ independent subproblems, each of which estimates the channel between the BS and one specific RIS element. In Steps 7-17, the subproblems are solved respectively using the coordinate descent method. As described in Steps 10-15, the outer iterations are run until a well-fit solution is found or the number of outer iterations reaches $I_{\rm max}$. In Step 13, we refine the $m$-th channel coefficients in the inner iterations. Finally, the algorithm will find an estimate of the quasi-static BS-RIS channel.

\begin{algorithm}[!tp]
{
\renewcommand{\algorithmicrequire}{\textbf{Input:}}
\renewcommand\algorithmicensure {\textbf{Output:} }
\caption{Proposed coordinate descent-based channel estimation algorithm}
\label{alg:Framework}
\begin{algorithmic}[1]
\REQUIRE
The received pilots $\left\{{\bar y}_{m_1,m_2,n}|\left(m,m'\right)\in {\mathcal S}\right\}$, the transmitted power $P_{\rm BS}$, the termination threshold $\epsilon$, the maximal number of outer iterations $I_{\rm max}$.

\ENSURE
Estimated coefficients $\left\{{\hat g}_{m,n}|1\le \!m\le M,1\le n\le\right.$ $\left. N\right\}$ of the BS-RIS channel.

\FOR{$m_1=1:L$}
\FOR{$m_2=1:M,m_2\neq m_1$}
\STATE {${\bar {\bf y}}_{m_1,m_2}^T \triangleq \left[{\bar y}_{m_1,m_2,1},{\bar y}_{m_1,m_2,2},\cdots,{\bar y}_{m_1,m_2,N+1}\right]$.}
\STATE {$\left[{\hat s}_{m_1,m_2}~~a_{m_1,m_2,1}~~a_{m_1,m_2,2}~~\cdots~~a_{m_1,m_2,N}\right]=\frac{1}{\sqrt{P_{\rm BS}\left(N+1\right)}}{\bar {\bf y}}_{m_1,m_2}^T{\bf F}_{N+1}^H$.}
\ENDFOR
\ENDFOR

\FOR{$n=1:N$}
\STATE {Initialize ${\hat g}^{(0)}_{1,n},{\hat g}^{(0)}_{2,n},\cdots,{\hat g}^{(0)}_{M,n}$ according to (\ref{equ:init_1})-(\ref{equ:init_2}).}
\STATE {$i=0$.}
\WHILE{$f_n\left({\hat g}^{(i)}_{1,n},\cdots,{\hat g}^{(i)}_{M,n}\right)>\epsilon$ \AND $i< I_{\rm max}$}
\STATE {$i\leftarrow i+1$.}
\FOR{$m=1:M$}
\STATE {Calculate ${\hat g}^{(i)}_{m,n}$ according to (\ref{equ:update_gmn_est}).}
\ENDFOR
\ENDWHILE
\STATE {${\hat g}_{m,n}={\hat g}_{m,n}^{(i)},1\le m\le M$.}
\ENDFOR

\RETURN{$\left\{{\hat g}_{m,n}|1\le m\le M,1\le n\le N\right\}$}

\end{algorithmic}
}
\end{algorithm}

\section{Mobile RIS-UE and BS-UE Channel Estimation}

Given the estimate of the quasi-static BS-RIS channel, we can estimate the mobile RIS-UE channel $\left\{{\bf f}_k|1\le k\le K\right\}$ and the BS-UE channel $\left\{{\bf h}_k|1\le k\le K\right\}$. Since the mobile channels are low-dimensional, they can be estimated with the conventional uplink pilot transmission scheme and an LS-based algorithm.

\subsection{Uplink pilot transmission}$~$

To estimate the mobile RIS-UE and BS-UE channels, we follow the conventional uplink pilot transmission scheme \cite{RISCE_Rayleigh_MVU}. The uplink pilot transmission frame consists of $\tau_0$ sub-frames, and each sub-frame lasts for $K$ time slots. In the $t$-th sub-frame ($t=1,2,\cdots,N+1$), the reflection coefficient vector at the RIS is ${\tilde {\boldsymbol \phi}}_t \in {\mathbb C}^{N\times 1}$. The reflection coefficient vector is randomly generated by ${\tilde {\boldsymbol \phi}}_t\left(n\right)=e^{j\omega_{t,n}}$, $t=1,2,\cdots,\tau_0$, where the random phase $\omega_{t,n}$ is uniformly distributed in $\left[0,2\pi\right)$, $n=1,2,\cdots,N$. During the $K$ time slots in a sub-frame, the UEs transmit uplink pilot sequences, i.e., ${\bf x}_k \in {\mathbb C}^{K\times 1}$, $k=1,2,\cdots,K$. In order to distinguish the pilots from different UEs, we assign orthogonal pilot sequences to different UEs, i.e.,
\begin{equation}
{\bf x}_{k_1}^H{\bf x}_{k_2}=\left\{\begin{matrix}KP_{\rm UE}, & k_1=k_2,\\0, & k_1\neq k_2,\end{matrix}\right.
\end{equation}
with $P_{\rm UE}$ denoting the transmitted power of each UE.

In the $t$-th sub-frame, based on (\ref{model:RIS_signal_model_equivalent}), we write the multi-slot pilot transmission model by
\begin{equation}\label{model:UE_uplink_pilot_transmission}
{\bf Y}_t = \sum\limits_{k=1}^{K}\left[{\bf G}{\rm diag}\left({\bf f}_k\right){\tilde {\boldsymbol \phi}}_t + {\bf h}_k\right]{\bf x}_k^T + {\bf N}_t,
\end{equation}
where ${\bf Y}_t\in {\mathbb C}^{M\times K}$ is the matrix of received pilots at the BS. Each column of ${\bf Y}_t$ is the received pilots in a single time slot. ${\bf N}_t\in {\mathbb C}^{M\times K}$ is the noise.

Then, by right multiplying the conjugate of the pilot sequences, we can distinguish the channels of different UEs:
\begin{equation}\label{model:phase_2_model_t}
\begin{aligned}
{\tilde {\bf y}}_{k,t} &= \frac{1}{KP_{\rm UE}}{\bf Y}_t{\bf x}_k^* \\
&= \frac{1}{KP_{\rm UE}}\sum\limits_{k'=1}^{K}\left[{\bf G}{\rm diag}\left({\bf f}_{k'}\right){\tilde {\boldsymbol \phi}}_t+{\bf h}_{k'} \right]{\bf x}_{k'}^T{\bf x}_k^* + \frac{{\bf N}_t{\bf x}_k^*}{KP_{\rm UE}}\\
&= \left[{\bf G}{\rm diag}\left({\tilde {\boldsymbol \phi}}_t\right){\bf f}_{k}+{\bf h}_{k} \right] + {\tilde {\bf n}}_{k,t}\\
&= {\bf A}_t\left[\begin{matrix}{\bf f}_{k}\\{\bf h}_{k}\end{matrix}\right] + {\tilde {\bf n}}_{k,t},~~k=1,2,\cdots,K,
\end{aligned}
\end{equation}
where ${\tilde {\bf n}}_{k,t}=\frac{{\bf N}_t{\bf x}_k^*}{KP_{\rm UE}}$, and
\begin{equation}\label{equ:definition_of_matrix_A_t}
{\bf A}_t\triangleq\left[\begin{matrix} {\bf G}{\rm diag}\left({\tilde {\boldsymbol \phi}}_t\right)&{\bf I}_{M}\end{matrix}\right].
\end{equation}

In (\ref{model:phase_2_model_t}), ${\tilde {\bf y}}_{k,t} \in {\mathbb C}^{M\times 1}$ is the equivalent received pilot for the $k$-th UE, $\left[{\bf f}_{k}^T,{\bf h}_{k}^T\right]^T \in {\mathbb C}^{\left(M+N\right)\times 1}$ is the vector of all coefficients of the mobile channels of the $k$-th UE. Since the number of received pilots must be no smaller than the dimension of the mobile channels, i.e., $\tau_0M\ge M+N$, we have $\tau_0\ge \tau_{\rm min}=\lceil \frac{M+N}{M} \rceil$. Then, by collecting the equivalent received pilots in $\tau_0$ frames, we have
\begin{equation}\label{model:phase_2_model_collected}
{\tilde {\bf y}}_{k} = {\bf A}\left[\begin{matrix}{\bf f}_{k}\\{\bf h}_{k}\end{matrix}\right] + {\tilde {\bf n}}_k, ~~ k=1,2,\cdots,K,
\end{equation}
where ${\tilde {\bf y}}_{k}\triangleq\left[{\tilde {\bf y}}_{k,1}^T,{\tilde {\bf y}}_{k,2}^T,\cdots,{\tilde {\bf y}}_{k,\tau_0}^T\right]^T$, ${\bf A}\triangleq \left[{\tilde {\bf A}}_{1}^T,{\tilde {\bf A}}_{2}^T,\right.$ $\left.\cdots,{\tilde {\bf A}}_{\tau_0}^T\right]^T$, and ${\tilde {\bf n}}_k\triangleq \left[{\tilde {\bf n}}_{k,1}^T,{\tilde {\bf n}}_{k,2}^T,\cdots,{\tilde {\bf n}}_{k,\tau_0}^T\right]^T$.

\subsection{LS-based channel estimation algorithm}$~$

In (\ref{model:phase_2_model_collected}), ${\bf A}$ is determined by the exact quasi-static channel ${\bf G}$ and the series of reflection coefficients $\left\{{\tilde {\boldsymbol \phi}}_t\Big|1\le t\le\tau_0\right\}$. The reflection coefficients $\left\{{\tilde {\boldsymbol \phi}}_t\Big|1\le t\le\tau_0\right\}$ are predesigned, while the exact value of $\bf G$ is not known to us. So we can only use ${\hat {\bf G}}$ which is estimated in the quasi-static channel estimation. Similar to (\ref{equ:definition_of_matrix_A_t}), we define ${\hat {\bf A}}$ based on ${\hat {\bf G}}$
\begin{equation}
{\hat {\bf A}}\triangleq\left[\begin{matrix}{\hat {\bf G}}{\rm diag}\left({\tilde {\boldsymbol \phi}}_1\right) & {\bf I}_{M}\\ \vdots & \vdots \\ {\hat {\bf G}}{\rm diag}\left({\tilde {\boldsymbol \phi}}_{\tau_0}\right) & {\bf I}_{M}\end{matrix}\right].
\end{equation}

Finally, we have the least square (LS) estimate of the mobile channels
\begin{equation}\label{equ:LS_channel_estimation}
\left[\begin{matrix}{\hat {\bf f}}_{k} \\ {\hat {\bf h}}_{k}\end{matrix}\right] = {\hat {\bf A}}^{\dagger}{\tilde {\bf y}}_{k}=\left({\hat {\bf A}}^H{\hat {\bf A}}\right)^{-1}{\hat {\bf A}}^H{\tilde {\bf y}}_{k},~~k=1,2,\cdots,K,
\end{equation}
where ${\hat {\bf f}}_{k}$ and ${\hat {\bf h}}_{k}$ are the estimates of the RIS-UE channel and the BS-UE channel of the $k$-th UE, respectively.

\section{Pilot Overhead and Computational Complexity}

In the quasi-static channel estimation, there are $\left(N+1\right)$ sub-frames, each of which consists of $L$ time slots. Taking the least value $L=2$, the associated pilot overhead is $\tau_1=\left(N+1\right)L=2\left(N+1\right)$. In the mobile channel estimation, there are $\tau_{\rm min}=\lceil \frac{M+N}{M} \rceil$ sub-frames, each of which consists of $K$ time slots, so the pilot overhead for the mobile channel estimation is $\tau_2=K\lceil\frac{M+N}{M} \rceil$.

In the proposed two-time channel estimation framework, the quasi-static BS-RIS channel is estimated in a large timescale. As a result, we need to average the pilot overhead of large-timescale channel estimation depending on how frequently the BS-RIS channel is estimated. Specifically, let $T_L$ and $T_S$ denote the channel coherence time of the large-timescale channel and that of the the small-timescale channel, respectively, and $T_L=\alpha T_S$ with $\alpha>>1$. During a time period of $T_S$, the average pilot overhead is calculated by $\tau=\frac{T_S}{T_L}\cdot\tau_1+\tau_2=\frac{2\left(N+1\right)}{\alpha}+K\lceil\frac{M+N}{M} \rceil$. The pilot overhead of the proposed two-timescale channel estimation is lower than $\left(M+N\right){K}$ in \cite{RISCE_Rayleigh_MVU}, and lower than $K+N+\max\left\{K-1,\lceil\frac{\left(K-1\right)N}{M}\rceil\right\}$ in \cite{RISCE_Rayleigh_multiuser_reduced} when $\alpha>2$.

In the quasi-static channel estimation, the computation complexity mainly lies in Step 4 and Step 14 of {\bf Algorithm 1}. In Step 4, we can use fast Fourier transformation (FFT) with the computational complexity of ${\mathcal O}\left(N\log N\right)$. In Step 13, it takes ${\mathcal O}\left(M\right)$ calculations when $m\le L$, and ${\mathcal O}\left(L\right)$ when $m>L$, so the computational complexity of Steps 12-14 is ${\mathcal O}\left(ML\right)$. Considering the number of iterations, the computational complexity of {\bf Algorithm 1} is ${\mathcal O}\left(MLN\log N + MLNI_{\rm max}\right)$. In the mobile channel estimation, the computational complexity of is determined by that of the LS channel estimation in (\ref{equ:LS_channel_estimation}), which is ${\mathcal O}\left(\left(M+N\right)^3+\left(M+N\right)^2K\right)$.

\section{Simulations}

\subsection{Simulation setup}

\begin{figure}[tp!]
\begin{center}
\includegraphics[width=0.6\columnwidth, height=0.45\columnwidth]{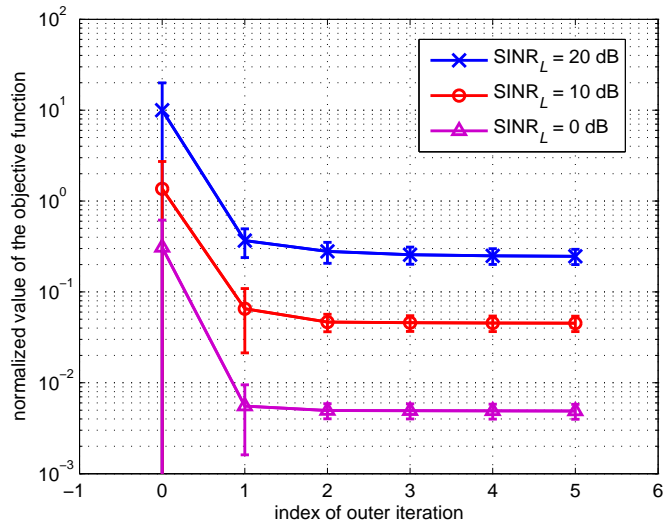}
\end{center}
\caption{The convergence of the coordinate descent-based channel estimation algorithm.}
\label{fig1}
\end{figure}

In our simulations, $M=32$, $N=64$, $K=8$. The large-scale fading is modeled as $\rho_{h_k}=\rho_{0}\left(\frac{d_{h_k}}{d_0}\right)^{-\alpha_h}$, $\rho_{g}=\rho_{0}\left(\frac{d_{g}}{d_0}\right)^{-\alpha_g}$, $\rho_{f_k}=\rho_{0}\left(\frac{d_{f_k}}{d_0}\right)^{-\alpha_f}$, where $\rho_{0}=-20$ dB is the large-scale fading factor at the reference distance of $d_0=1$ m, and we set $\alpha_h=2.2$, $\alpha_g=2.1$ and $\alpha_f=4.2$, which are also adopted in the simulations in \cite{RISCE_Rayleigh_multiuser_reduced}. The BS-RIS distance is $d_{g}=20$ m, the BS-UE distance is $d_{h_k}=30$ m, and the RIS-UE distance is $d_{f_k}=20$ m. The transmit power of the BS and the UEs is $P_{\rm BS}=10$ W and $P_{\rm UE}=1$ W, respectively. The proposed two-timescale channel estimation consists of the large-timescale channel estimation and the small-timescale channel estimation, which are influenced by different noises. For the dual-link pilot transmission in the large-timescale channel estimation, we define the signal-to-interference-plus-noise ratio by
\begin{equation}
{\rm SINR}_L\triangleq\frac{P_{\rm BS}\rho_{g}^2}{\sigma_{i}^2+\sigma_{n}^2},
\end{equation}
where the after-mitigation self-interference level is set to be 2 times as strong as the receiver noise, i.e., $\sigma_{\rm SI}^2=2\sigma_{n}^2$ \cite{FD_SI_Cancellation1}. While for the uplink pilot transmission in the small-timescale channel estimation, we define the signal-to-noise ratio by
\begin{equation}
{\rm SNR}_S=\frac{P_{\rm UE}\rho_{r}\rho_{g}}{\sigma_{n}^2}.
\end{equation}

\subsection{Simulation results}

\begin{figure}[tp!]
\begin{center}
\includegraphics[width=0.6\columnwidth, height=0.45\columnwidth]{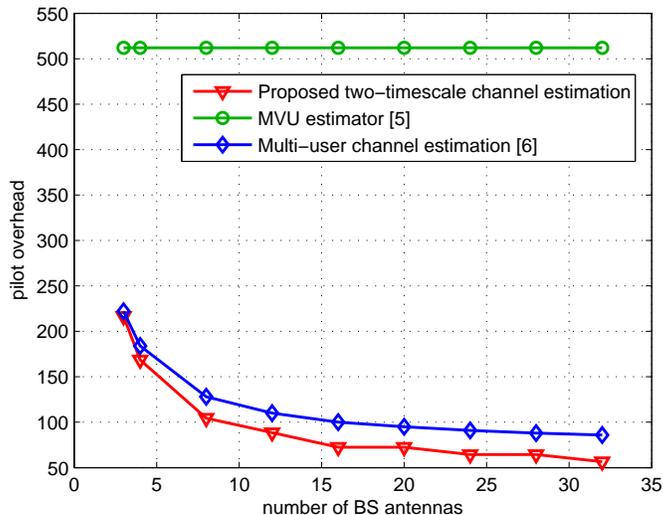}
\end{center}
\caption{The pilot overhead of different channel estimation schemes.}
\label{fig1}
\end{figure}

First of all, we examine the convergence of the proposed large-timescale channel estimation algorithm by simulations. We randomly generate 100 channels and run the entire channel estimation procedure for each of them. In each channel estimation, we need to solve $N=64$ subproblems (\ref{problem:multivariate_optimization}). For each subproblem, we can evaluate the normalized value of the objective function by
\begin{equation}
{\bar f}_n^{(i)}\triangleq \frac{f_n\left({\hat g}^{(i)}_{1,n},\cdots,{\hat g}^{(i)}_{M,n}\right)}{{\mathbb E}\left\{f_n\left(0,\cdots,0\right)\right\}},~~i=1,\cdots,I_{\rm max},
\end{equation}
so we can record how the value of ${\bar f}_n^{(i)}$ decreases with $i$ for the $100N=6,400$ independent subproblems. In Fig. 3, we show the means and standard deviations of ${\bar f}_n^{(i)}$ by the curves of error bars, under ${\rm SINR}_L=0~dB$, ${\rm SINR}_L=10~dB$ and ${\rm SINR}_L=20~dB$. As shown in Fig. 3, it takes no more than 5 outer iterations for the coordinate descent-based algorithm to converge to a stable mean value with a low standard deviation. Therefore, we set the maximum number of outer iterations $I_{\rm max}=5$ for {\bf Algorithm 1}.

Then, we compare the pilot overhead of different channel estimation schemes in Fig. 4. We can see that the pilot overhead of the proposed two-timescale channel estimation framework is lower than the multi-user channel estimation method in \cite{RISCE_Rayleigh_multiuser_reduced}, and significantly lower than the minimum variance unbiased (MVU) estimator \cite{RISCE_Rayleigh_MVU}, by leveraging the two-timescale channel property.

In Fig. 5-6, we investigate the normalized mean square error (NMSE) performance for the proposed two-timescale channel estimation framework. Most existing channel estimation methods only estimate the BS-UE direct channel $\left\{{\bf h}_k|1\le k\le K\right\}$ and the BS-RIS-UE cascaded channel $\left\{{\bf C}_k|1\le k\le K\right\}$. As a result, to compare the NMSE performance with the existing cascaded channel estimation methods, we should also calculate the estimate of the cascaded channel $\left\{{\bf C}_k|1\le k\le K\right\}$ based on our estimates of ${\bf G}$ and $\left\{{\bf f}_k|1\le k\le K\right\}$ according to ${\hat {\bf C}}_k={\hat {\bf G}}{\rm diag}\left({\hat {\bf f}}_k\right)$, $1\le k\le K$. The NMSE of the cascaded channel is defined by
\begin{equation}
{\rm NMSE}_{C}\triangleq\frac{{\mathbb E}\left\{\sum\limits_{k=1}^{K}\left\|{\hat {\bf C}}_k-{\bf C}_k\right\|_{F}^2\right\}}{{\mathbb E}\left\{\sum\limits_{k=1}^{K}\left\|{\bf C}_k\right\|_{F}^2\right\}},
\end{equation}
while the NMSE of the BS-UE channel is
\begin{equation}
{\rm NMSE}_{h}\triangleq\frac{{\mathbb E}\left\{\sum\limits_{k=1}^{K}\left\|{\hat {\bf h}}_k-{\bf h}_k\right\|_2^2\right\}}{{\mathbb E}\left\{\sum\limits_{k=1}^{K}\left\|{\bf h}_k\right\|_2^2\right\}}.
\end{equation}

\begin{figure}[tp!]
\begin{center}
\includegraphics[width=0.6\columnwidth, height=0.45\columnwidth]{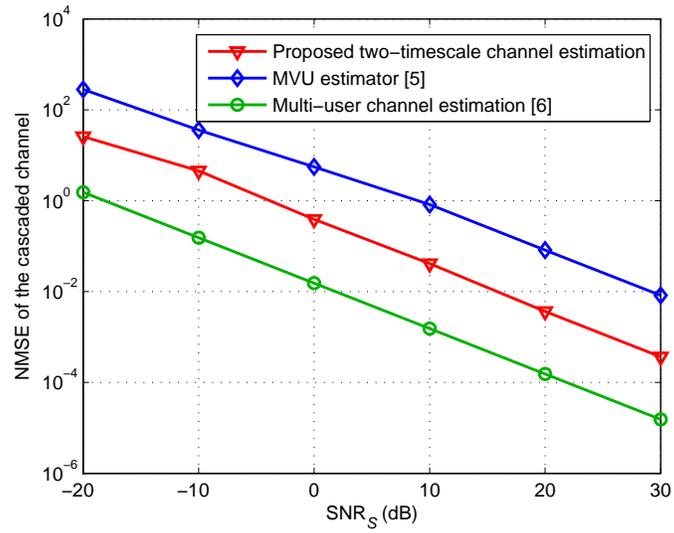}
\end{center}
\caption{The NMSE for the cascaded channel against uplink SNR.\protect\\~~~~~~~~~~~~~~~~~~~~~~~~~~~~~~~~~~~~~~~~~~~}
\label{fig1}
\end{figure}

\begin{figure}[tp!]
\begin{center}
\includegraphics[width=0.6\columnwidth, height=0.45\columnwidth]{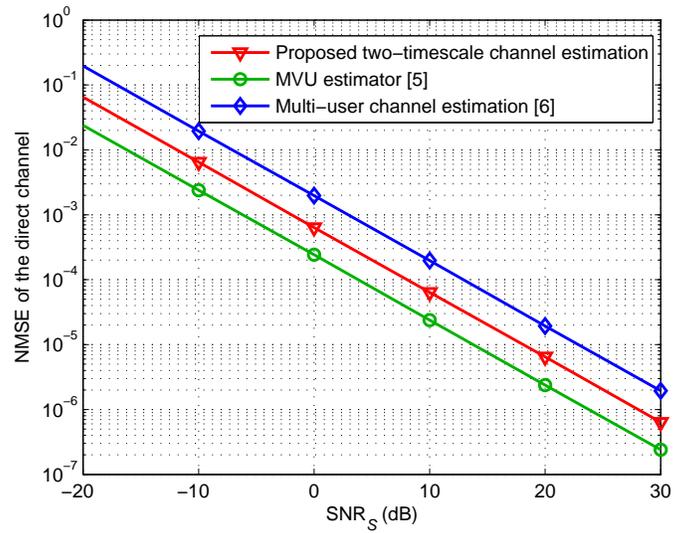}
\end{center}
\caption{The NMSE for the direct channel against uplink SNR.}
\label{fig1}
\end{figure}

Fig. 5 shows the NMSE of the cascaded channel against the SNR, while Fig. 6 shows the NMSE of the BS-UE channel against the SNR. The proposed channel estimation method can achieve lower NMSE than that in \cite{RISCE_Rayleigh_multiuser_reduced}. The MVU channel estimation \cite{RISCE_Rayleigh_MVU} is more accurate, but it is mainly because the pilot overhead is $32$ times as high as that of our proposed method.

Fig. 7 is the sum-rate performance comparison. We adopt a cross entropy optimization-based precoding scheme similar to that in \cite{CEO_precoding} to jointly optimize the precoding matrix at the BS and the reflection coefficient vector at the RIS. The precoding is conducted based on the CSI estimated from different schemes. The sum-rate based on perfect CSI is also adopted as an upper bound for comparison. The result is consistent with Figs. 5-6, that the proposed channel estimation method can outperform the method in \cite{RISCE_Rayleigh_multiuser_reduced}.

\begin{figure}[tp!]
\begin{center}
\includegraphics[width=0.6\columnwidth, height=0.45\columnwidth]{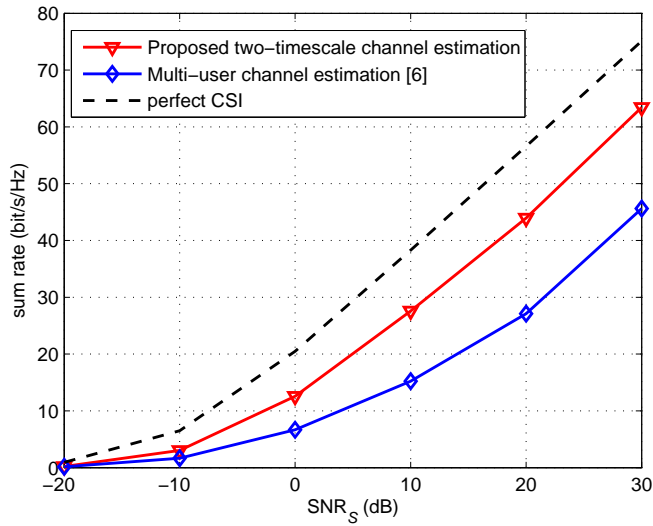}
\end{center}
\caption{The sum-rate comparison against uplink SNR.}
\label{fig1}
\end{figure}

\section{Conclusions}

In this paper, we propose a two-timescale channel estimation framework for the RIS-aided wireless communication systems. The key idea of the framework is to exploit the two-timescale property, which means that the BS-RIS channel is high-dimensional but quasi-static, while the RIS-UE channel and the BS-UE channel are mobile but low-dimensional. We reveal that the pilot overhead can be significantly reduced by exploiting the two-timescale channel property. For the quasi-static BS-RIS channel, the average pilot overhead can be reduced from a long-term perspective since it is not estimated frequently. For the mobile RIS-UE and BS-UE channels that have to be frequently estimated in a small timescale, their dimension is smaller than that of the cascaded channel, so the pilot overhead can be significantly reduced. Simulation results show that the proposed two-timescale channel estimation framework can achieve accurate channel estimation with low pilot overhead. For future research, it is worth to extend the proposed two-timescale channel estimation to adapt to different channel models and scenarios.

\end{document}